\documentclass{article}
\usepackage{amsmath}
\usepackage{cite}
\usepackage{graphicx}
\usepackage{dcolumn}

\begin{document}

\date{}
\title{Just another conditionally-solvable non-relativistic quantum-mechanical model}
\author{Francisco M. Fern\'{a}ndez\thanks{%
fernande@quimica.unlp.edu.ar} \\
%EndAName
INIFTA, DQT, Sucursal 4, C. C. 16, \\
1900 La Plata, Argentina}
\maketitle

\begin{abstract}
We show that a perturbed Coulomb problem discussed recently is conditionally
solvable. We obtain the exact eigenvalues and eigenfunctions and compare the
former with eigenvalues calculated by means of a numerical method. We
discuss the meaning of the numbers that determine the exact solutions which
arise from the Frobenius (power-series) method.
\end{abstract}

\section{Introduction}

\label{sec:intro}

In a paper published recently, Omugbe et al\cite{OOEOA21} proposed a
non-relativistic quantum-mechanical model with a central-field potential
that they called perturbed Coulomb potential. It is the sum of a Coulomb
potential and a truncated Coulomb potential\cite{F20c} (and references
therein). The truncated Coulomb potential is an example of
conditionally-solvable problems that have been studied since long ago (see,
for example, \cite{T16,CDW00} and the references therein). The solutions of
some conditionally-solvable quantum-mechanical problems have been used in
several fields of theoretical physics but in many cases they were grossly
misinterpreted\cite{AF20,F21} (and references therein).

As mentioned above, it is well known that the Schr\"{o}dinger equation with
the truncated Coulomb potential is conditionally solvable\cite{F20c}. In
this paper we investigate if the sum of a Coulomb potential and a truncated
Coulomb potential also leads to a conditionally solvable Schr\"{o}dinger
equation. In section~\ref{sec:model} we briefly discuss the model and the
derivation of dimensionless eigenvalue equations. In section~\ref
{sec:conditional_sol} we study the Scr\"{o}dinger equation by means of the
Frobenius (power-series) method\cite{AF20,F21}. Finally, in section~\ref
{sec:conclusions} we summarize the main results and draw conclusions.

\section{The model}

\label{sec:model}

The Hamiltonian operator for the model is
\begin{equation}
H=-\frac{\hbar ^{2}}{2m}\nabla ^{2}+V(r),\;V(r)=-\frac{V_{1}}{r}+\frac{V_{2}%
}{r+r_{0}},  \label{eq:H}
\end{equation}
where $V_{1}>0$ and $r_{0}>0$. For simplicity, we assume that $V_{1}>V_{2}$
that leads to an infinite number of bound states.

It is convenient to convert the Schr\"{o}dinger equation into a
dimensionless eigenvalue equation by means of the transformations $\left(
x,y,z\right) \rightarrow \left( L\tilde{x},L\tilde{y},L\tilde{z}\right) $, $%
\nabla ^{2}\rightarrow L^{-2}\tilde{\nabla}^{2}$, where $L$ is a suitable
unit of length\cite{F20}. The dimensionless Hamiltonian operator is
\begin{equation}
\tilde{H}=\frac{mL^{2}}{\hbar ^{2}}H=-\frac{1}{2}\tilde{\nabla}^{2}-\frac{%
mLV_{1}}{\hbar ^{2}\tilde{r}}+\frac{mLV_{2}}{\hbar ^{2}\left( \tilde{r}+%
\tilde{r}_{0}\right) },  \label{eq:H_dim_L}
\end{equation}
where $\tilde{r}=r/L$ and $\tilde{r}_{0}=r_{0}/L$. Note that the unit of
energy is $\hbar ^{2}/\left( mL^{2}\right) $.

Case I: the unit of length $L=r_{0}$ leads to the unit of energy $\hbar
^{2}/\left( mr_{0}^{2}\right) $ and
\begin{equation}
\tilde{H}=-\frac{1}{2}\tilde{\nabla}^{2}-\frac{a}{\tilde{r}}+\frac{b}{\left(
\tilde{r}+1\right) },\;a=\frac{mr_{0}V_{1}}{\hbar ^{2}},\;b=\frac{mr_{0}V_{2}%
}{\hbar ^{2}}.  \label{eq:H_dim_r0}
\end{equation}

Case II: the unit of length $L=\hbar ^{2}/\left( mV_{1}\right) $ leads to
the unit of energy $mV_{1}^{2}/\hbar ^{2}$ and
\begin{equation}
\tilde{H}=-\frac{1}{2}\tilde{\nabla}^{2}-\frac{1}{\tilde{r}}+\frac{V_{2}}{%
V_{1}\left( \tilde{r}+\tilde{r}_{0}\right) }.  \label{eq:H_dim_V1}
\end{equation}

Case III: the unit of length $L=\hbar ^{2}/\left( mV_{2}\right) $ leads to
the unit of energy $mV_{2}^{2}/\hbar ^{2}$ and
\begin{equation}
\tilde{H}=-\frac{1}{2}\tilde{\nabla}^{2}-\frac{V_{1}}{V_{2}\tilde{r}}+\frac{1%
}{\left( \tilde{r}+\tilde{r}_{0}\right) }.  \label{eq:H_dim_V2}
\end{equation}
One chooses the units that are most suitable for one's purposes. In what
follows we focus on the dimensionless Hamiltonian operator (\ref{eq:H_dim_r0}%
) and omit the tilde on the dimensionless quantities from now on.

The Schr\"{o}dinger equation with a central-field potential is separable in
spherical coordinates $r$, $\theta $, $\phi $ and the eigenfunctions can be
written as $\psi _{\nu l}(r,\theta ,\phi )=R_{\nu l}(r)Y_{l}^{m}(\theta
,\phi )$, where $Y_{l}^{m}(\theta ,\phi )$ are the spherical harmonics and $%
l=0,1,\ldots $, $m=0,\pm 1,\pm 2,\ldots ,\pm l$ are the angular-momentum
quantum numbers. The radial quantum number $\nu =0,1,\ldots $ is the number
of nodes of the radial part $R_{\nu l}(r)$ of the solution in the interval $%
0<r<\infty $. The eigenvalues of the Schr\"{o}dinger equation are commonly
written $E_{\nu l}$ in order to stress their dependence on the quantum
numbers $\nu $ and $l$.

\section{Conditional solution}

\label{sec:conditional_sol}

The radial part of the Schr\"{o}dinger equation for Case I can be written as
\begin{equation}
-\frac{1}{2}f^{\prime \prime }(r)+\left[ \frac{l(l+1)}{2r^{2}}-\frac{a}{r}+%
\frac{b}{r+1}\right] f(r)=Ef(r),  \label{eq:radial_eq}
\end{equation}
where $f(r)=r^{-1}R(r)$.

If we look for a solution of the form
\begin{equation}
f(r)=e^{-\alpha r}r^{l+1}\sum_{j=0}^{\infty }c_{j}r^{j},\;\alpha =\sqrt{-2E},
\label{eq:f(r)}
\end{equation}
then the expansion coefficients $c_{j}$ satisfy the three-term recurrence
relation
\begin{eqnarray}
c_{j+2} &=&A_{j}c_{j+1}+B_{j}c_{j},\;j=0,1,\ldots ,  \nonumber \\
A_{j} &=&-\frac{2a-2\alpha \left( j+l+2\right) +\left( j+1\right) \left[
j+2\left( l+1\right) \right] }{\left( j+2\right) \left( j+2l+3\right) },
\nonumber \\
B_{j} &=&-2\frac{a-\alpha \left( j+l+1\right) -b}{\left( j+2\right) \left(
j+2l+3\right) }.  \label{eq:TTRR}
\end{eqnarray}
In order to obtain exact polynomial solutions we require that $c_{n}\neq 0$
and $c_{n+1}=c_{n+2}=0$, $n=0,1,\ldots $. In this way $c_{j}=0$ for all $j>n$%
. These conditions require that $B_{n}=0$ from which we obtain an expression
for $\alpha $:
\begin{equation}
\alpha =\frac{a-b}{n+l+1}.  \label{eq:alpha}
\end{equation}
One may think that in this way we have obtained the energy eigenvalues of
the problem
\begin{equation}
E=-\frac{\alpha ^{2}}{2}=-\frac{\left( a-b\right) ^{2}}{2\left( n+l+1\right)
^{2}},  \label{eq:E(a,b,n,l)}
\end{equation}
but it is not the case because the condition $c_{n+1}(l,a,b)=0$ links the
values of the model parameters $a$ and $b$, the quantum number $l$ and the
polynomial order $n$.

Because of equation (\ref{eq:alpha}) the quantities $A_{j}$ and $B_{j}$
become
\begin{eqnarray}
A_{j} &=&\frac{2a\left( j-n+1\right) -2b\left( j+l+2\right) -\left(
j+1\right) \left( j+2\left( l+1\right) \right) \left( l+n+1\right) }{\left(
j+2\right) \left( j+2l+3\right) \left( l+n+1\right) },  \nonumber \\
B_{j} &=&\frac{2\left( a-b\right) \left( j-n\right) }{\left( j+2\right)
\left( j+2l+3\right) \left( l+n+1\right) },  \label{eq:Aj,Bj_n}
\end{eqnarray}
and the infinite series in equation (\ref{eq:f(r)}) reduces to the
polynomial
\begin{equation}
p^{(n)}(r)=\sum_{j=0}^{n}c_{j}^{(n)}r^{j},  \label{eq:p^(n)(r)}
\end{equation}
where, without loss of generality, we arbitrarily set $c_{0}=1$ from now on.
In what follows we discuss some particular cases.

When $n=0$ we have $c_{1}^{(0)}=-b/(l+1)=0$ that leads to $b=0$ and the
energies of the pure Coulomb model $E_{0l}=-a^{2}/\left[ 2\left( l+1\right)
^{2}\right] $ for $\nu =n=0$.

When $n=1$ the coefficient
\begin{equation}
c_{2}^{(1)}=\frac{b\left( a+\left( b+l+2\right) \left( l+1\right) \right) }{%
\left( l+1\right) \left( l+2\right) \left( 2l+3\right) },  \label{eq:c_2^(1)}
\end{equation}
is a quadratic function of $b$ with roots
\begin{equation}
b^{(1,1)}=0,\;b^{(1,2)}=-\frac{a+\left( l+1\right) \left( l+2\right) }{l+1}.
\label{eq:b^(1,i)}
\end{equation}
When $b=b^{(1,1)}=0$ we obtain the energies $E_{1l}=-a^{2}/\left[ 2\left(
l+2\right) ^{2}\right] $ of the pure Coulomb model and the polynomial
\begin{equation}
p^{(1,1)}(r)=1-\frac{a}{(l+1)(l+2)}r.  \label{eq:p^(1,1)(r)}
\end{equation}
Note that in this case $n=\nu =1$ because the exact eigenfunction exhibits
one radial node. On the other hand, when $b=b^{(1,2)}$ we have
\begin{equation}
p^{(1,2)}(r)=1+r,  \label{eq:p^(1,2)(r)}
\end{equation}
that is nodeless ($\nu =0$) and, consequently, $n=1$ cannot be interpreted
as being the radial quantum number.

When $n=2$, $c_{3}^{(2)}$ is a cubic function of $b$. Since the roots, which
can be easily obtained, are rather cumbersome to be shown here we will
discuss just one of them $b=b^{(2,1)}=0$ that also appears in this case. We
obtain the energies $E_{2l}=-a^{2}/\left[ 2\left( l+3\right) ^{2}\right] $
of the pure Coulomb model and the exact eigenfunctions with the polynomial
\begin{equation}
p^{(2,1)}(r)=1-\frac{2a}{\left( l+1\right) \left( l+3\right) }r+\frac{2a^{2}%
}{\left( l+1\right) \left( l+3\right) ^{2}\left( 2l+3\right) }r^{2},
\label{eq:p^(2,1)(r)}
\end{equation}
that exhibits two nodes. Once again, we see that in the pure Coulomb case $%
n=\nu $ is the radial quantum number. For the other two roots $\nu =n-1=1$
and $\nu =n-2=0$ and $n$ cannot be considered to be the radial quantum
number.

In general, $c_{n+1}^{(n)}=0$ exhibits $n+1$ roots $b^{(n,i)}$, $%
i=1,2,\ldots ,n+1$, with always the root $b^{(n,1)}=0$. We conclude that the
power-series ansatz (\ref{eq:f(r)}) only yields the exact result for the
trivial case $b=0$ and conditional solutions for the other roots. With these
roots we obtain the exact energies
\begin{equation}
E_{l}^{(n,i)}=--\frac{\left[ a-b^{(n,i)}\right] ^{2}}{2\left( n+l+1\right)
^{2}},  \label{eq:E_l^(n,i)}
\end{equation}
that are useless if we do not organize and connect them properly\cite{F21}.
This approach does not provide exact solutions for $b>0$ because $%
b^{(n,i)}\leq 0$ for all $n$. Strictly speaking, we should write $%
b^{(n,i)}(l)$ and $c_{j}^{(n,i)}(l)$ to emphasize the dependence of these
quantities on the rotational quantum number $l$.

In order to test present analytical expressions we resort to the Riccati-Pad%
\'{e} method (RPM) that commonly yields highly accurate eigenvalues of
separable Schr\"{o}dinger equations\cite{FMT89a} and choose $n=1$ and $l=0$
as a particular case. For example, when $a=1$ then $b^{(1,2)}=-3$ and $%
E_{0}^{(1,2)}=-2$. The RPM yields $E_{00}=-2$, $E_{10}=-0.8399328077$, $%
E_{20}=-0.4725478081$ and $E_{30}=-0.3042665976$ for the four lowest
eigenvalues. When $a=2$ then $b^{(1,2)}=-4$ and $E_{0}^{(1,2)}=-4.5$. In
this case the RPM gives us $E_{00}=-4.5$, $E_{10}=-1.819915414$,\ $%
E_{20}=-1.023806204$ and $E_{30}=-0.464083021$. It is well known that the
RPM yields the exact polynomial solutions correctly. In the examples just
considered it gives us the exact value for $E_{00}$ in agreement with the
power-series approach developed above. The agreement between these two
completely different procedures is a clear confirmation that present
analytical expressions are correct.

Figure~\ref{Fig:Enu0} shows $E_{0}^{(n,i)}$ for $a=1$ and some values of $n$
and $i$ as well as the lowest RPM eigenvalues for a range of values of $b$.
We appreciate that the RPM curves $E_{\nu 0}(b)$, $\nu =0,1,2,3$, pass
through some of the eigenvalues $E_{0}^{(n,i)}$ thus revealing the true
meaning of the latter as eigenvalues of a given Hamiltonian operator. Note
that the RPM eigenvalues satisfy the Hellmann-Feynman theorem\cite{G32,F39}
\begin{equation}
\frac{\partial E}{\partial b}=\left\langle \frac{1}{r+1}\right\rangle >0.
\label{eq:HFT}
\end{equation}

Figure~\ref{Fig:E0L} shows similar results for $\nu =0$ and $l=0,1,2$. As
expected $E_{\nu l^{\prime }}>E_{\nu l}$ for $l^{\prime }>l$.

\section{Conclusions}

\label{sec:conclusions}

\begin{figure}[tbp]
\begin{center}
\includegraphics[width=9cm]{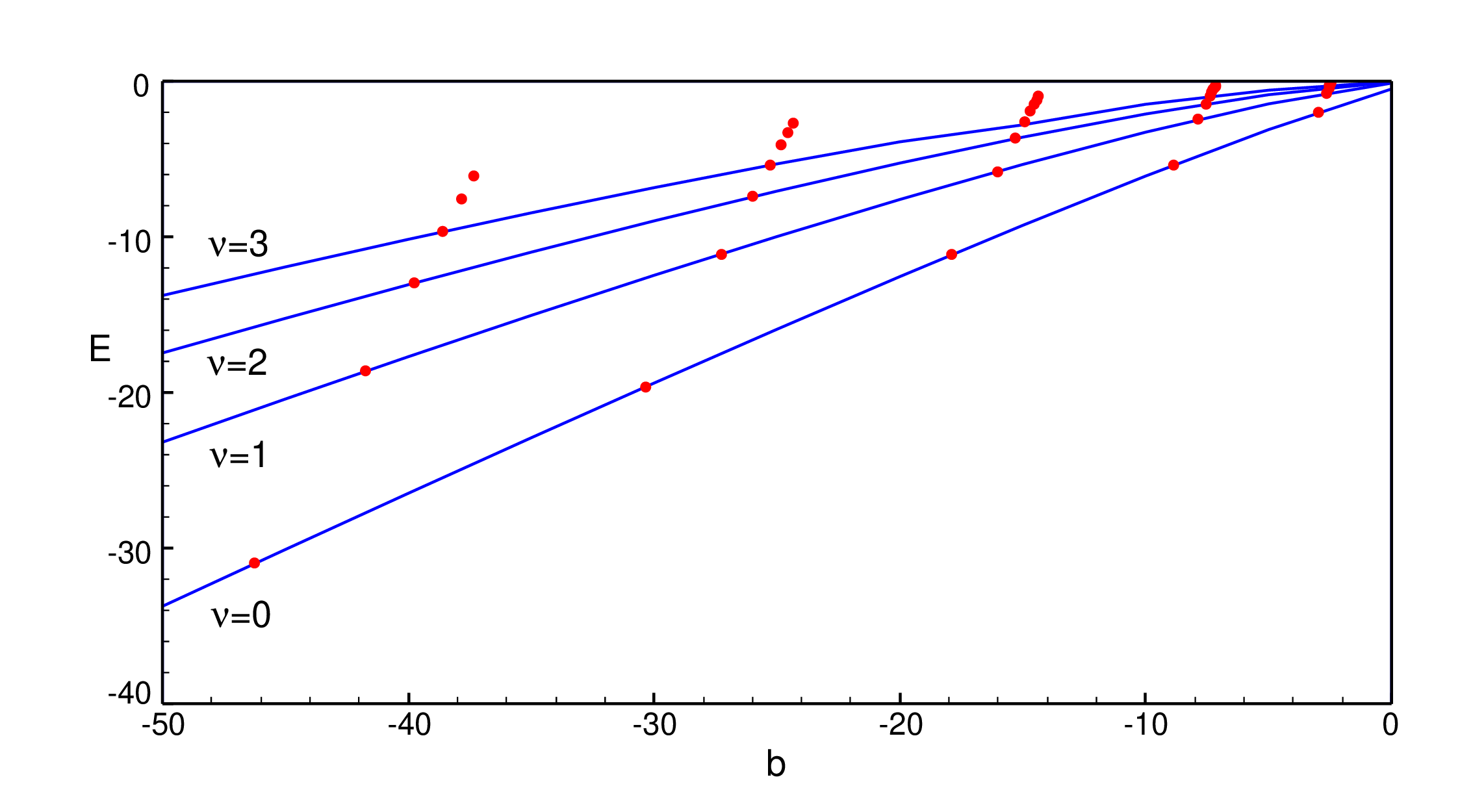}
\end{center}
\caption{Exact eigenvalues $E_{0}^{(n,i)}$ coming from polynomial solutions
(red circles) and RPM eigenvalues $E_{\nu 0}$ (blue lines)}
\label{Fig:Enu0}
\end{figure}

\begin{figure}[tbp]
\begin{center}
\includegraphics[width=9cm]{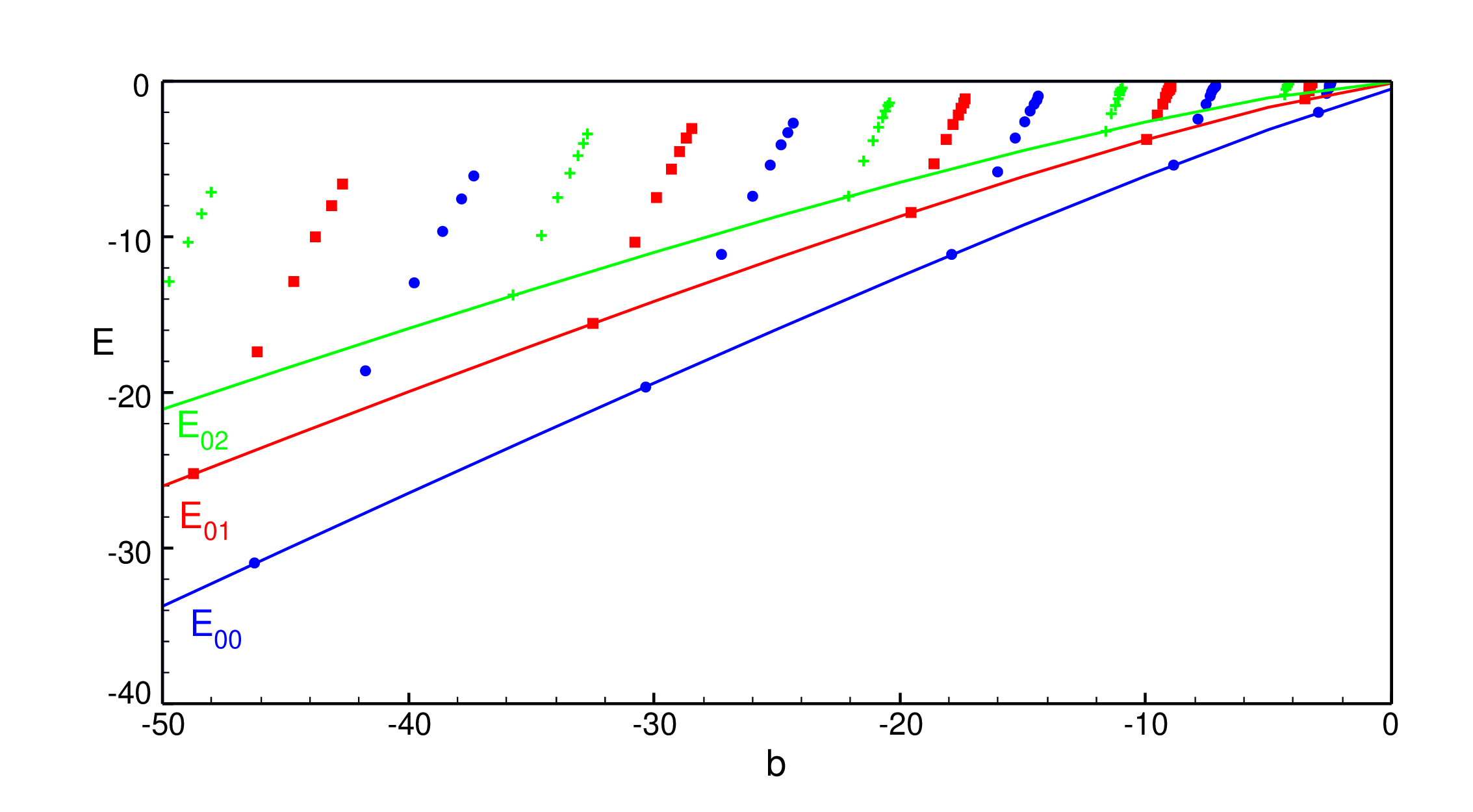}
\end{center}
\caption{Exact eigenvalues $E_{l}^{(n,i)}$ coming from polynomial
solutions (blue circles ($l=0$), red squares ($l=1$), green
crosses ($l=2$)) and RPM eigenvalues $E_{\nu 0}$ (blue ($l=0$),
red ($l=1$) and green ($l=2$) lines)} \label{Fig:E0L}
\end{figure}

We have shown that the Schr\"{o}dinger equation with the Hamiltonian
operator (\ref{eq:H}) is conditionally solvable. For some values of $E$ ($%
=E_{l}^{(n,i)}$) and $b$ ($=b_{l}^{(n,i)}$) we obtain exact eigenfunctions
with a polynomial factor $p_{l}^{(n,i)}(r)$ of degree $n$. The quantities $%
E_{l}^{(n,i)}$ are not eigenvalues of a given physical model but eigenvalues
of models with different values of $b$. The number $n$ is the order of the
polynomial but it is not the radial quantum number because it is not the
number of radial nodes of the wavefunction for all values of $i$. In fact, $%
n\geq \nu $, where $\nu $ is the actual radial quantum number. The
eigenvalues $E_{l}^{(n,i)}$ are useless if they are not properly organized
and connected as shown in present figures \ref{Fig:Enu0} and \ref{Fig:E0L}
(see also\cite{F20c,AF20,F21}. The solutions of conditionally solvable
problems like the one studied here have been grossly misinterpreted by many
authors as discussed elsewere\cite{AF20,F21}.

\end{document}